\documentclass[aps,prl,twocolumn,amsmath]{revtex4-1} 
\pdfoutput=1
\usepackage{feynmp, tensor}
\usepackage{ulem}

\usepackage{color,graphicx}
\usepackage{bm}

\usepackage{amsmath}
\usepackage{amsfonts}
\usepackage{amssymb}

\usepackage{subfigure}
\usepackage[latin1]{inputenc}

\newcommand{\Balpha}{\boldsymbol\alpha}
\newcommand{\Bbeta}{\boldsymbol\beta}
\newcommand{\Bgamma}{\boldsymbol\gamma}

\renewcommand{\-}{\,-\,}

\newcommand{\bk}{\mathbf{k}}
\newcommand{\bq}{\mathbf{q}}

\newcommand{\SRO}{Sr$_2$RuO$_4$ }
\newcommand{\SROdot}{Sr$_2$RuO$_4$}

\newcommand{\be}{\begin{equation}}
\newcommand{\ee}{\end{equation}}

\newcommand{\bea}{\begin{equation}\begin{aligned}}
\newcommand{\eea}{\end{aligned}\end{equation}}

\let\oldmarginpar\marginpar
\renewcommand\marginpar[1]{\-\oldmarginpar[\raggedleft\tiny #1]%
{\raggedright\tiny #1}}

\newcommand{\subfigimg}[3][,]{%
  \setbox1=\hbox{\includegraphics[#1]{#3}}
  \leavevmode\rlap{\usebox1}
  \rlap{\hspace*{-5pt}\raisebox{\dimexpr\ht1-2\baselineskip}{#2}}
  \phantom{\usebox1}
}

\begin{document}


\title{Large Chern Number and Edge Currents in \SRO}
\author{Thomas Scaffidi}
\affiliation{Rudolf Peierls Centre for Theoretical Physics, Oxford OX1 3NP, United Kingdom}

\author{Steven H. Simon}
\affiliation{Rudolf Peierls Centre for Theoretical Physics, Oxford OX1 3NP, United Kingdom}

\date{\today}

\begin{abstract}
We show from a weak coupling microscopic calculation that the most favored chiral superconducting order parameter in \SRO has Chern number $|C|=7$.
The two dominant components of this order parameter are given by $\sin(3 k_x)  + i \sin(3 k_y)$ and $\sin(k_x) \cos(k_y) + i \sin(k_y) \cos(k_x)$ and lie in the same irreducible representation $E_u$ of the tetragonal point group as the usually assumed gap function $\sin(k_x) + i \sin(k_y)$.
While the latter gap function leads to $C=1$,  the two former lead to $C =-7$, which is also allowed for an $E_u$ gap function since the tetragonal symmetry only fixes $C$ modulo 4.
Since it was shown that the edge currents of a $|C|>1$ superconductor vanish exactly in the continuum limit, and can be strongly reduced on the lattice, this form of order parameter could help resolve the conflict between experimental observation of time-reversal symmetry breaking and yet the absence of observed edge currents in \SROdot.
%
\end{abstract}

\maketitle

\SRO is a layered perovskite material exhibiting a transition at 1.5 K to an unconventional superconducting phase.
There is a lot of experimental evidence in favour of an odd-parity, possibly topological, superconducting phase \cite{maeno1994superconductivity,RevModPhys.75.657,JPSJ.81.011009,0953-8984-21-16-164210,0034-4885-75-4-042501}.
These topological superconductors come in two kinds: chiral and helical.
Chiral superconductors break time-reversal symmetry, have a $\mathbb{Z}$ topological number (called hereby the Chern number $C$ and defined below) and {\it can} exhibit edge currents while helical superconductors are time-reversal symmetric, have a $\mathbb{Z}_2$ topological number and can only exhibit time-reversed pairs of helicity currents.
Majorana states in chiral superconductors could be used for topological quantum information processing\cite{RevModPhys.80.1083}.


Evidence for time-reversal symmetry breaking in \SRO was given by muon spin relaxation \cite{luke1998time} and optical Kerr effect \cite{PhysRevLett.97.167002} experiments. 
These experiments therefore point towards a chiral superconductor.
The order parameter (OP) of a triplet superconductor is given by a three-dimensional vector $\vec{d}(\bk)$ \cite{RevModPhys.75.657}.
For a tetragonal crystal like \SROdot, this OP should transform according to a given representation of $D_{4h}$.
Among the odd-parity irreducible representations of $D_{4h}$, the only one corresponding to a chiral state is $E_u$, for which the order parameter should be given by $d_z=h_x+i \ h_y$ (or $d_z=h_x-i \ h_y$ for the opposite chirality) where $h_{x,y}$ stands for any function of momentum that transforms in the same way as $\sin(k_{x,y})$ under the symmetry operations of $D_{4h}$.

The simplest example of such a gap function is given by 
\be
d_{z,0}^{\nu}(\bk) \equiv \sin(k_x) + i \sin(k_y) \ \forall \ \nu,
\ee
where $\nu$ is the band index. 
This OP has been used as a prevailing assumption in the field.
In this case, in analogy with superfluid $^3$He-A, the superconducting state is supposed to be driven by ferromagnetic fluctuations on the fairly isotropic $\boldsymbol\gamma$ band, which is therefore the dominant band in this scenario.
The two other bands, called $\boldsymbol\alpha$ and $\Bbeta$, are then merely spectators.

\begin{figure}
  \centering
  \begin{tabular}{@{}p{0.45\linewidth}@{\quad}p{0.45\linewidth}@{}}
    \subfigimg[width=\linewidth]{a)}{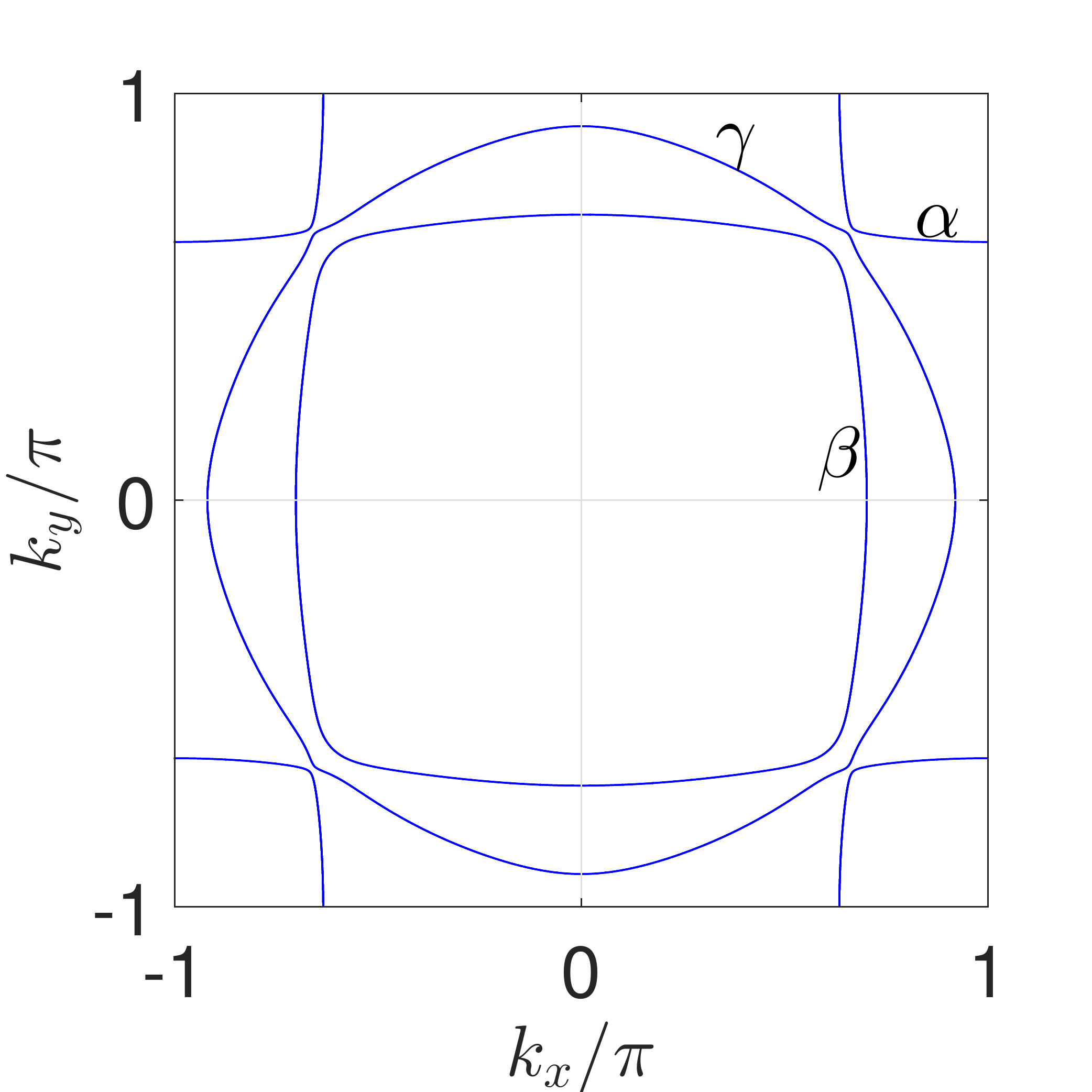} &
    \subfigimg[width=\linewidth]{b)}{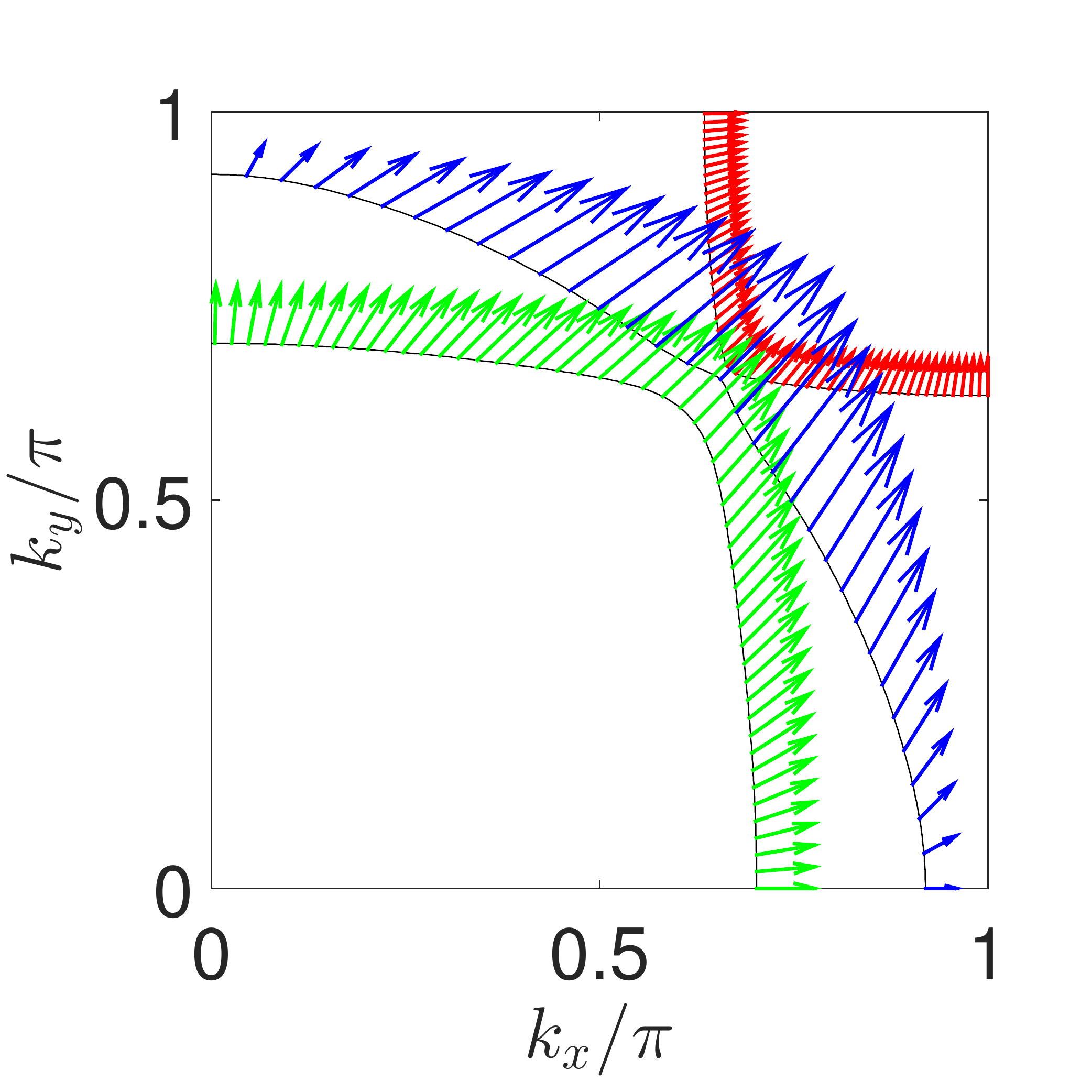} \\
    \subfigimg[width=\linewidth]{c)}{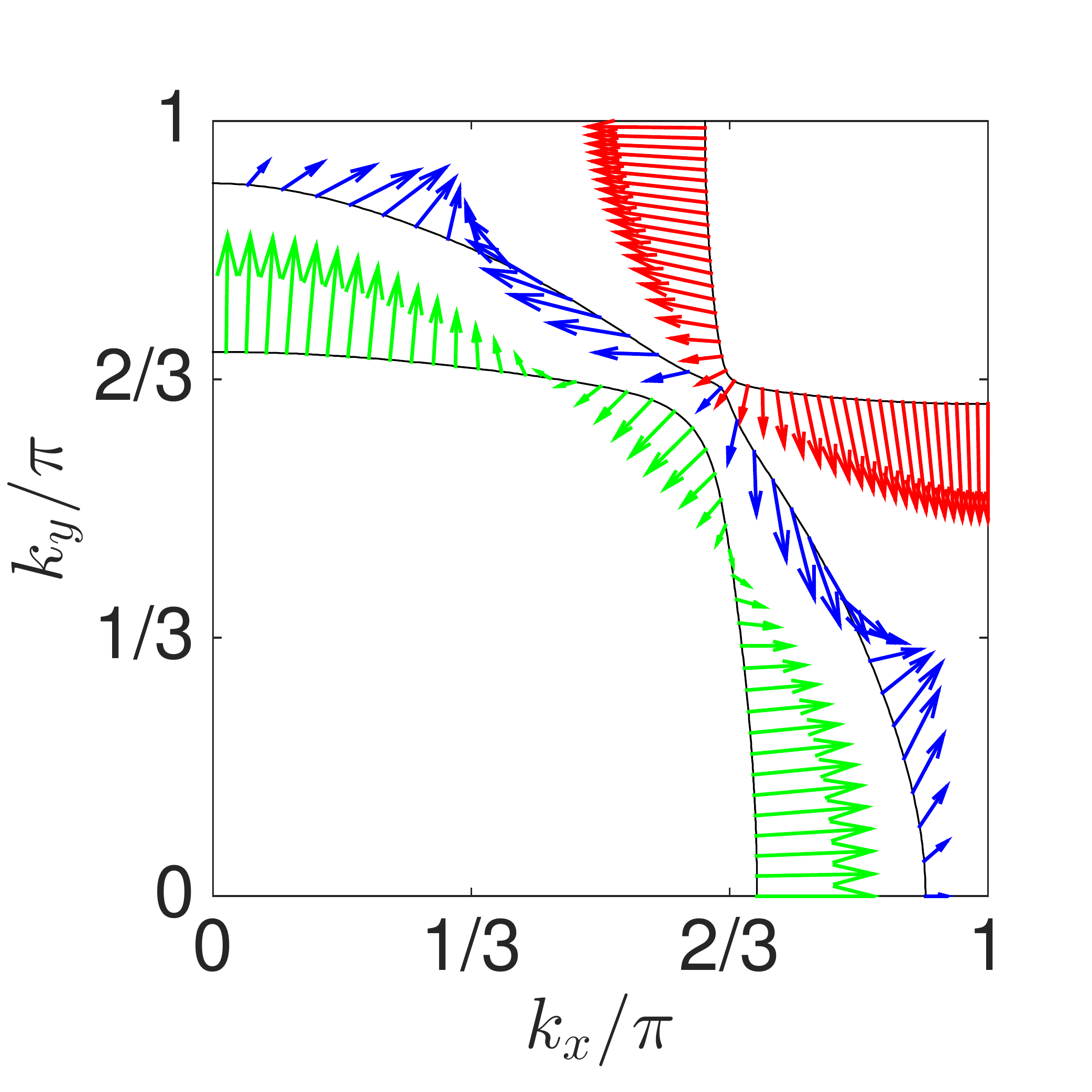} &
    \subfigimg[width=\linewidth]{d)}{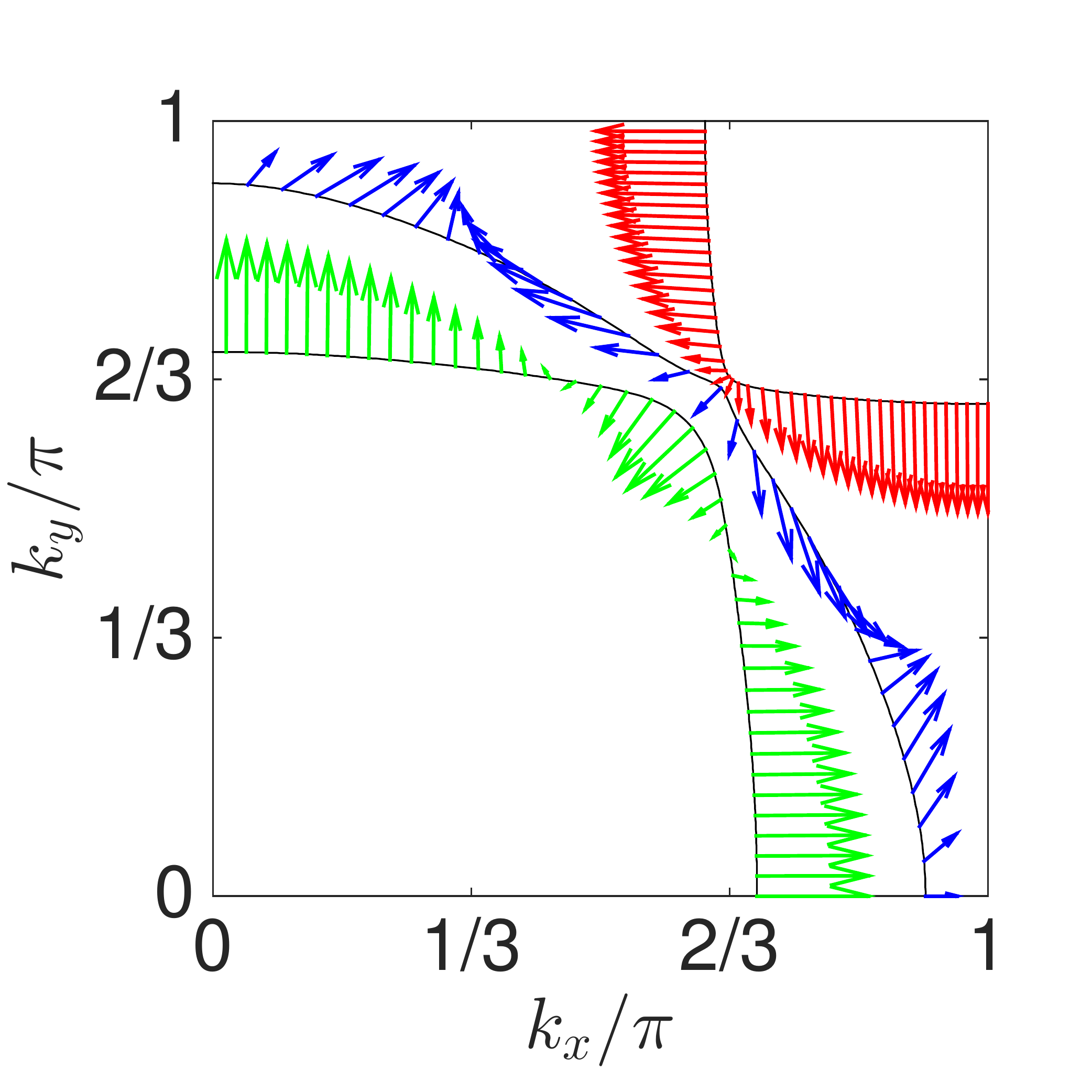} 
  \end{tabular}
\caption{\label{Gaps}
(a) Fermi surfaces for the tight-binding model from Ref.~\cite{PhysRevB.89.220510}. (b) Order parameter $d_{z,0}^{\nu}(\bk)$. The $x$ and $y$ components of the arrows give the real and imaginary part of $d_z$, respectively. The units are arbitrary. (c),(d) Same plot for $d_{z,\text{RG}}^{\nu}(\bk)$, $d_{z,\text{Fit}}^{\nu}(\bk)$, respectively.  Even though the gap has deep minima, it remains finite at all $\bk$. 
}
\end{figure}

Since there are three bands at the Fermi level (see Fig.~\ref{Gaps}(a)), the Chern number $C$ is given by the sum of the Chern number of each band $C_{\nu}$.
The Chern number is given by the winding of the complex phase of $d_z$ around the Fermi surface (FS) of a given band, or it is equivalently given by the skyrmion number of the BdG Hamiltonian\cite{1999JETPL..70..609V}:
\be
C_{\nu} = \frac1{4\pi} \int d\bk \ \hat{H}_{\nu} \cdot \left( \partial_{k_x}\hat{H}_{\nu} \times \partial_{k_y} \hat{H}_{\nu} \right)
\ee
where $\vec{H}=\{\operatorname{Re}[d_z(\bk)], -\operatorname{Im}[d_z(\bk)], E(\bk) - \mu \}$, $\hat{H}=\vec{H}/|\vec{H}|$, $E(\bk)$ is the band dispersion and $\mu$ is the chemical potential.
Considering $d_{z,0}$ as shown in Fig.~\ref{Gaps}(b), it is easy to see that $C_{\nu}=+1$ for a FS centered at $(0,0)$ (i.e. a particle band) and $C_{\nu}=-1$ for a FS centered at $(\pi,\pi)$ (i.e. a hole band).
Since there are two particle bands ($\boldsymbol\beta$ and $\boldsymbol\gamma$) and one hole band ($\boldsymbol\alpha$) in \SROdot, the total Chern number in this case is $C=1$.



The issue with this scenario is that a chiral superconductor with $C=1$ should exhibit a nonzero total orbital angular momentum and edge currents, which have been elusive so far despite intense scrutiny\cite{PhysRevB.76.014526,PhysRevB.81.214501,PhysRevB.89.144504}.
Spontaneous angular momentum and currents in chiral superfluids have been studied extensively\cite{Ishikawa01061977,PhysRevB.21.980,kita1998angular,PhysRevB.64.054514,PhysRevB.69.184511,PhysRevB.84.214509} and it was
confirmed recently that, for $C = 1$, in both the continuum and the lattice OP $d_{z,0}$ case, these currents are quite inevitable\cite{2014arXiv1409.7459T,2014arXiv1409.8638V,PhysRevB.90.224519,PhysRevB.91.094507}.
The apparent contradiction between measurements of time-reversal symmetry breaking and the absence of edge currents has been a long-standing puzzle about \SROdot\cite{0953-8984-21-16-164210}.

The dominant $\Bgamma$ scenario was challenged by a renormalization group (RG) calculation\cite{PhysRevB.81.224505,PhysRevB.83.094518,PhysRevB.88.064505} that showed that, in the weak-coupling limit, the quasi-one-dimensional (1D) $\Balpha$ and $\Bbeta$ bands are actually driving superconductivity through antiferromagnetic fluctuations caused by the nesting of their FSs\cite{PhysRevLett.82.4324,PhysRevLett.85.4586,doi:10.1143/JPSJ.69.3505,PhysRevB.62.R14641,PhysRevB.63.060506,PhysRevLett.105.136401,PhysRevB.86.064525}.
The gap was therefore thought to be dominant on these two bands whose total Chern number is zero for an OP given by $d_{z,0}$, thereby making \SRO a topologically trivial superconductor.
STM data showed that these bands have a gap amplitude in accordance with BCS theory given the value of $T_c$, thus supporting the idea that the main gap is on the $\alpha$ and $\beta$ bands\cite{PhysRevB.88.134521}.
The problem with this scenario is that, from thermodynamic data, it is believed that the gap should be of similar size on all three bands\cite{JPSJ.69.572,PhysRevLett.92.047002,PhysRevB.88.134521}, and that therefore $\Bgamma$ should have a sizable gap which must lead to a nontrivial topology and presumably sizable edge currents.
Futhermore, while the Chern numbers of $\Balpha$ and $\Bbeta$ are opposite in the case of $d_{z,0}$, this is not true in general, and it is in particular not true for the type of order parameter favored by the nesting of $\Balpha$ and $\Bbeta$, as we will show later on.


In a previous work, we extended the aforementioned RG technique to include inter-band coupling and spin-orbit coupling at the microscopic level\cite{PhysRevB.89.220510}. 
The inclusion of these effects was shown to be crucial since it enabled us to obtain a similarly-sized gap on all three bands without any fine tuning, in agreement with thermodynamic data and in contrast to previous results.
Depending on the ratio of Hund's coupling $J$ to Hubbard interation $U$, this calculation could either favour a chiral state in the $E_u$ representation, or a helical state in the $A_{1u}$ representation. 
Because of the evidence of time-reversal symmetry breaking, we will focus on the former case in this paper\footnote{Note that our discussion also applies to the favored helical state in the $A_{1u}$ representation favored by RG in some subset of the parameter range \cite{PhysRevB.89.220510}, as long as Chern number is replaced by mirror Chern number\cite{PhysRevLett.111.087002}}.
The gap function we obtain in the $E_u$ representation $d_{z,\text{RG}}^{\nu}(\bk)$ has a highly non-trivial momentum dependence (see Fig.~\ref{Gaps}(c)), indicative of pairing with a range longer than nearest neighbors.






The main result of this work is the following: instead of having $C=+1$ for each particle band, like for $d_{z,0}$ and for a continuum $p_x+ip_y$ state, the OP $d_{z,\text{RG}}$ has a Chern number of $-3$ for the particle bands $\Bbeta$ and $\Bgamma$, as seen in Fig.~\ref{Gaps}(c).
This is allowed by symmetry, since being in the $E_u$ representation fixes $C$ to be 1, but only modulo 4.
Adding $C_{\Balpha} = -1$ (which has a different value from $\Bbeta$ and $\Bgamma$ since it is a hole-like band instead of a particle-like band), this leads to a total Chern number of $-7$.
This is a dramatic change compared to the continuum case and this shows that, when lattice effects are strong, it can be misleading to have continuum OPs in mind.

Before discussing the experimental implications of this result, let us first give an intuitive understanding of the source of this longer range pairing.
Generically, the real (imaginary) part of an OP in the $E_u$ representation, called $h_x$ ($h_y$), can be written as a linear combination of all possible harmonics $g_{x}(\bk)$ ($g_{y}(\bk)$) that transform under $D_{4h}$ in the same way as $\sin(k_x)$ ($\sin(k_y)$).
%
The most simple one is obviously $g_{x,1}(\bk)=\sin(k_x)$ and corresponds to nearest neighbor pairing.
We find that the non-trivial anisotropy of $ d^{\nu}_{z,\text{RG}}(\bk)$ originates from two longer range pairing components, $g_{x,2}(\bk) \equiv \sin(k_x) \cos(k_y)$ and $g_{x,3}(\bk)\equiv \sin(3 k_x)$, that are favored on all three orbitals.
These components are favored due to the presence of strong fluctuations at the nesting wavevectors $(\pm 2\pi/3, \pi)$ and $(\pm 2\pi/3, \pm 2\pi/3)$, respectively.


In the weak coupling limit, the effective interaction in the odd-parity superconducting channel generically takes the following form
\be
V(\bk,\bq) = - U^2 \chi(\bk-\bq)
\label{diagram}
\ee
where $\chi$ is the susceptibility and has maxima at the nesting wavevectors $\mathbf{Q}$.
The most favored superconducting OP $\Delta(\bk)$ is the eigenvector of $V(\bk,\bq) $ with the most negative eigenvalue.
In order to achieve a maximally negative eigenvalue, it is favorable to have 
\be
\arg\left[\Delta(\bk+\mathbf{Q})\right] = \arg\left[\Delta(\bk)\right]
\label{Cond}
\ee
where $\bk$ and $\bk+\mathbf{Q}$ both lie on the FS.
Depending on the value of $\mathbf{Q}$, this will favor certain gap functions over others.



As stated earlier, the driving force behind superconductivity are the strong fluctuations created by the nesting of the $\Balpha$ and $\Bbeta$ FSs.
These FSs are generated by the small hybridization of the $d_{xz}$ and $d_{yz}$ orbitals, whose unhybridized FSs are given by almost straight lines at $k_x = \pm k_F$ (resp. $k_y = \pm k_F$), with $k_F \simeq 2\pi/3$.
If we neglect hybridization for now and focus on $d_{xz}$, the nesting wavevectors are given by $(\pm 2 k_F, \pi)$ and the constraint can be rewritten as 
\be
\arg\left[\Delta(k_F-2 k_F,k_y+\pi)\right] = \arg\left[\Delta(k_F,k_y)\right].
\ee
For $\sin(k_x)$, these two values have opposite sign and it is therefore expected for this pairing to be strongly suppressed on the quasi-1D orbitals.
On the contrary, the gap function $\sin(k_x)\cos(k_y)$, corresponding to a second neighbor pairing, satisfies the above constraint and is expected to be favored\cite{PhysRevLett.105.136401}\footnote{Second neighbor pairing was already shown to lead to phases with $|C|>1$ in Refs.~\cite{PhysRevB.88.184513,PhysRevB.90.220511}}.

Now, once hybridization is taken into account, the nesting wavevector becomes $\mathbf{Q}=(\pm 2 k_F, \pm 2 k_F)\simeq(\mp 2\pi/3, \mp 2\pi/3)$, in accordance with neutron data\cite{PhysRevLett.83.3320}.
In this case, the condition from Eq.~\ref{Cond} is clearly satisfied by the function $\sin(3 k_x)$, corresponding to a pairing with a neighbor separated by three lattice constants along $[100]$\cite{2015arXiv150301352T}.


The argument given so far only applied to the quasi-1D orbitals.
Yet, thanks to spin-orbit coupling, inter-orbital hopping and inter-orbital interaction, superconductivity naturally arises on all the three bands, even though nesting originates from $\Balpha$ and $\Bbeta$\cite{PhysRevB.89.220510}.
We therefore expect $g_{2}(\bk)$ and $g_{3}(\bk)$ to be present along with $g_{1}(\bk)$ on the quasi-2D orbital $d_{xy}$ which dominantly contributes to $\Bgamma$ at the Fermi level.
The contribution of these gap functions to $d^{\nu}_{z,\text{RG}}(\bk)$ can be made explicit by using the following ansatz for the gap in orbital space
\bea
\Delta^a_{\text{Fit}}(\bk) &= \sum_{j=1,2,3} \Delta^a_{x,j} \ g_{x,j}(\bk) + i \ \Delta^a_{y,j} \ g_{y,j}(\bk) \\
g_{x,1}(\bk) &= \sin(k_x) \\
g_{x,2}(\bk) &= \sin(k_x) \cos(k_y) \\
g_{x,3}(\bk) &= \sin(3 k_x)
\label{GapOrbital}
\eea
where $a=xz,yz,xy$ is the orbital index, $g_{y,j}(k_x,k_y)=g_{x,j}(k_y,k_x)$ and  $\Delta^{zy}_{y,j} = \Delta^{zx}_{x,j}$; $\Delta^{zx}_{y,j} = \Delta^{zy}_{x,j} = 0 $; $\Delta^{xy}_{x,j} = \Delta^{xy}_{y,j} \ \forall \ j$.
In order to compare this ansatz with $d^{\nu}_{z,\text{RG}}(\bk)$, we apply to $\Delta^a_{\text{Fit}}(\bk)$ a momentum-dependent unitary transformation obtained by diagonalization of the spin-orbit-coupled hopping Hamiltonian given in Ref.~\cite{PhysRevB.89.220510}.
By doing so, we obtain the corresponding gap in band space, $d^{\nu}_{z,\text{Fit}}(\bk)$.
%
As seen in Fig.~\ref{Gaps}(c-d), we find that $d^{\nu}_{z,\text{Fit}}(\bk) \simeq d^{\nu}_{z,\text{RG}}(\bk)$ for the following parameters: $(\Delta^{zx}_{x,1},\Delta^{zx}_{x,2},\Delta^{zx}_{x,3})=(0,0.2,1.0)$ and $(\Delta^{xy}_{x,1},\Delta^{xy}_{x,2},\Delta^{xy}_{x,3})=(0.18,0.15,-0.3)$.

Possible experimental implications of a higher Chern number are now discussed\footnote{We note that the properties that depend only on the parity of $C$, like the number of Majorana bound states in a half-quantum vortex \cite{Jang14012011} or on a dislocation line linked with a vortex line \cite{PhysRevB.90.235123}, do not distinguish between $d_{z,0}$ and $d_{z,\text{RG}}$}.
First, $C$ gives the number of branches of chiral Majorana modes that can be found at sample edges and at dislocations with a Burgers vector whose component along [001] is non-zero\cite{2010arXiv1006.5454R,PhysRevB.90.235123}\footnote{Non-trivial 1D $Z_2$ weak topological indices $\boldsymbol\nu=(\nu_x,\nu_y,\nu_z)$ can also lead to Majorana modes at dislocations \cite{2010arXiv1006.5454R,PhysRevB.90.235123}. We find that, unlike the 2D topological index $C$, these invariants do not distinguish between $d_{z,0}$ and $d_{z,\text{RG}}$ since they are given in both cases by $\boldsymbol\nu=(1,1,0)$.}.
This could lead to specific signatures in tunneling measurements \cite{PhysRevLett.107.077003,doi:10.7566/JPSJ.83.074706,PhysRevB.88.134521} and edge state spectroscopy using angle-resolved photoemission spectroscopy.
These chiral Majorana modes lead to a quantization of the low temperature thermal Hall conductance, whose value is proportional to $C$ \cite{PhysRevB.61.10267,doi:10.7566/JPSJ.82.023602}:
\be
K_{xy} = \frac{C}{2} \frac{\pi^2 k_B^2 T}{6 \pi \hbar},
\label{ThermalHall}
\ee
where $k_B$ is Boltzmann constant and $T$ is the temperature.


We now discuss implications for edge currents in \SROdot.
Since charge is not conserved in a superconductor, the charge Hall conductance $G_{xy}$ is not universal and depends on the microscopic details, unlike $K_{xy}$.
%
In the continuum, due to rotational symmetry, there is only one possible OP for a given value of $C$: $d_z \propto (p_x + i p_y)^C$.
Taking advantage of this, it was shown that having edge currents and a total orbital angular momentum ``of order one'' is inevitable for a $|C|=1$ chiral superfluid in the continuum~\cite{2014arXiv1409.7459T,2014arXiv1409.8638V,PhysRevB.90.224519,PhysRevB.91.094507}\footnote{This can be seen from Eq.~\ref{K3} where the Fermi surface average vanishes identically for $|C|>1$. As shown in Ref.~\cite{PhysRevB.90.224519}, in the continuum, one has $v_{x}\propto\cos(\theta)$, $v_{y}\propto\sin(\theta)$, $h_{x}\propto\cos(C \theta)$ and $h_{y}\propto\sin(C \theta)$, where $\theta=\arctan(p_y/p_x)$. The Fermi surface average is then simply an integral over $\theta$, which vanishes for $|C|>1$.}. 
On the contrary, these two quantities were shown to vanish in the case of $|C|>1$ \cite{2014arXiv1409.7459T,2014arXiv1409.8638V,PhysRevB.90.224519,PhysRevB.91.094507} \footnote{As pointed out in Refs~\cite{PhysRevB.90.224519,PhysRevB.91.094507}, for smooth enough edge potential (which is not expected to be physically relevant), the edge currents are never suppressed for any non-zero Chern number.}.

\begin{table}
\caption{\label{k3tab}Chern numbers and Ginzburg-Landau coefficients (arbitrary units) for the two order parameters studied in this work }
\begin{ruledtabular}
\begin{tabular}{ccccccccc}
OP & $C_{\Balpha}$ & $C_{\Bbeta}$& $C_{\Bgamma}$& C &$k_{3,\Balpha}$ &  $k_{3,\Bbeta}$ & $k_{3,\Bgamma}$ & $\overline{k_3}$ \\
\hline\hline
$d_{z,0}$ & -1 & 1 & 1 & 1 & 0.50 & 0.99 & 1.14 & 1.0 \\
$d_{z,\text{RG}}$ & -1 & -3  & -3 & -7& -0.04 & 0.07 & -0.14 & -0.06
\end{tabular}
\end{ruledtabular}
\end{table}

When lattice effects cannot be neglected, like for \SROdot, there are lots of possible OPs for a given Chern number, and the aforementioned dichotomy present in the continuum breaks down.
In this case, the magnitude of edge currents can vary greatly from one OP to the other, even if they have the same Chern number.
%
In order to estimate the edge currents for the different OPs discussed in this work, we follow the Ginzburg-Landau (GL) calculation given in Ref.~\cite{PhysRevB.90.224519,PhysRevB.91.094507} (see also Refs.~\cite{RevModPhys.63.239,doi:10.1080/00018739400101475,PhysRevB.64.054514,PhysRevB.90.220511,PhysRevB.90.134521}).
In this theory, it can be shown that the current density coming from band $\nu$ is proportional to the following coefficient
\be
k_{3,\nu} \propto \langle h_{x,\nu}(\bk) h_{y,\nu}(\bk) v_{x,\nu}(\bk) v_{y,\nu}(\bk) \rangle_{\text{FS}_{\nu}}
\label{K3}
\ee
where $h_x$ and $h_y$ are the dimensionless real and imaginary part of the order parameter and $v_{x,y}$ are the Fermi velocity components and the average is over the FS.
The total current is proportional to the average of the $k_{3,\nu}$ coefficients weighted by the respective density of states at the Fermi level: $\overline{k_3} = (1/\rho) \times \sum_{\nu} \rho_{\nu} k_{3,\nu}$.
We note that, from Eq.~\ref{K3}, it is confirmed that the Chern number and the value of edge currents are not directly related for a lattice system.
Indeed, by applying to a given OP a rapid rotation of $\vec{h}$ over a small portion of the FS, it is possible to change the Chern number without changing $k_{3,\nu}$ significatively.
Such modification of the OP is not possible in the continuum because it breaks rotational symmetry.
In Table \ref{k3tab}, we give the values of $k_{3,\nu}$ and $\overline{k_3}$ for $d_{z,0}$ and $d_{z,\text{RG}}$.
We find that $\overline{k_3}$ is reduced by a factor of roughly 20 for $d_{z,\text{RG}}$ compared to $d_{z,0}$.

\begin{figure}[t!]
 \includegraphics[scale=0.3]{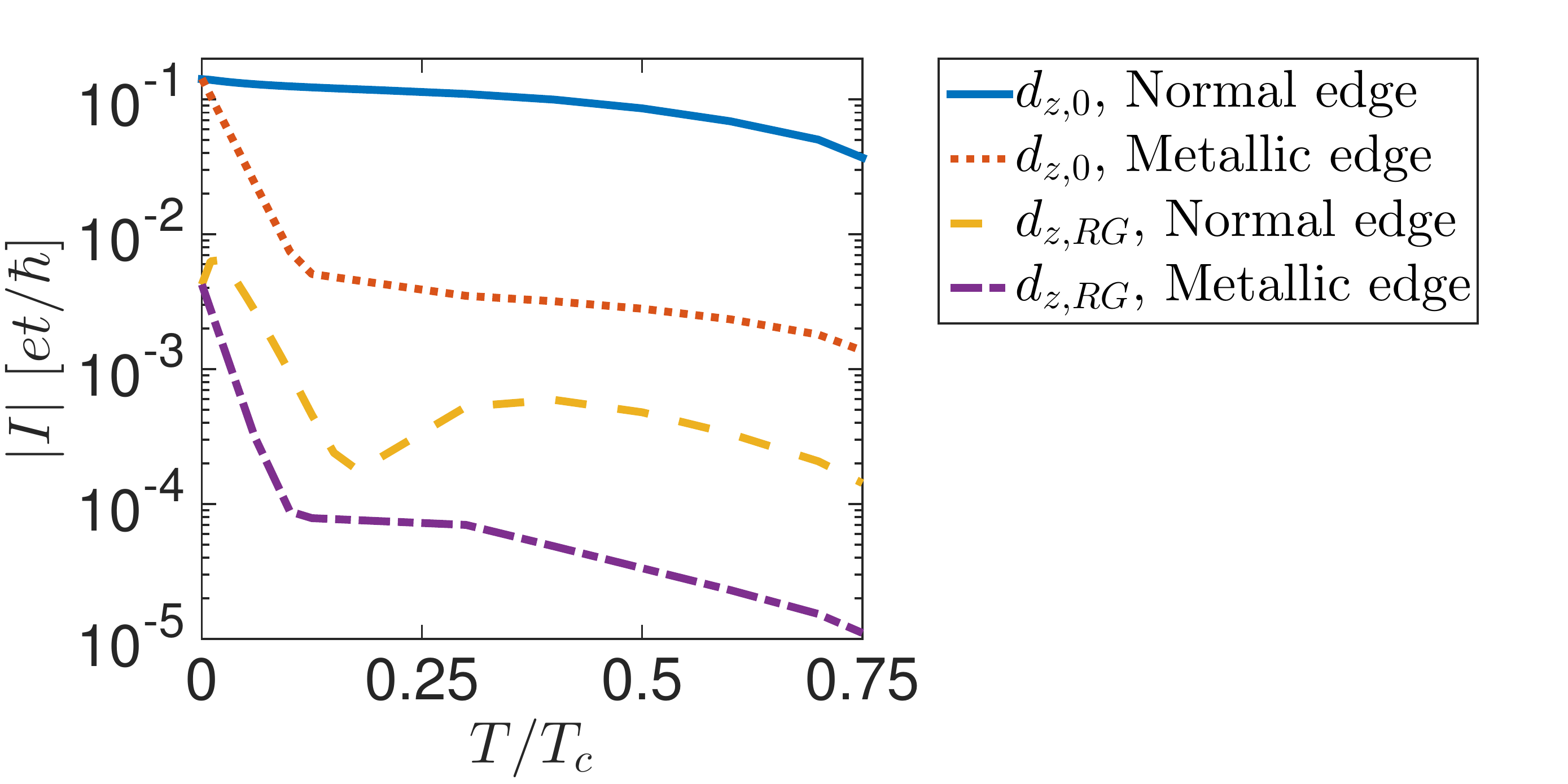}
\caption{Spontaneous currents $I_{0}$ and $I_{\text{RG}}$ for the gap functions $d_{z,0}$ and $d_{z,\text{RG}}$, respectively. 
These results were obtained from a BdG calculation in a cylinder geometry of the spin-orbit-coupled, three orbital hopping Hamiltonian studied in Ref.\cite{PhysRevB.89.220510}.
A superconducting region of width $L_S = 750$ sites was taken in which the gap takes a uniform value given by $T_c = 0.01t$.
In the metallic edge case, a region of width $L_M=500$ sites was added at the edge in which the gap is set to zero.
\label{BdG}
} 
\end{figure}

Since the gap $d_{z,\text{RG}}$ has deep minima, it is expected that finite temperature effects should lead to a large current reduction over a temperature scale set by this gap minima.
In order to estimate this effect for $d_{z,\text{RG}}$\footnote{The OPs are actually needed in orbital space for BdG calculations. We therefore used $\Delta^a_{\text{Fit}}(\bk)$ given in Eq.~\ref{GapOrbital} for the case of $d_{z,\text{RG}}$ and we used nearest-neighbor pairing on all three orbitals for the case of $d_{z,0}$.}, we perform a Bogoliubov-de Gennes (BdG) calculation in a cylinder geometry for the spin-orbit-coupled, three orbital hopping Hamiltonian studied in Ref.\cite{PhysRevB.89.220510}.
In Fig.~\ref{BdG}, we show the spontaneous currents $I_0$ and $I_{\text{RG}}$ for $d_{z,0}$ and $d_{z,\text{RG}}$, respectively.
We find that (1) at zero temperature, $I_{\text{RG}}$ is reduced by a factor of 30 compared to $I_0$, in overall agreement with the Ginzburg-Landau result; (2) unlike in the case of $d_{z,0}$, finite temperature effects generate a large drop in current in the case of $d_{z,\text{RG}}$.
We emphasize that this reduction should be very robust and appear both at edges and domain walls, since it comes from an intrinsic property of the bulk superconducting state.

There are other proposals for edge currents reduction \cite{PhysRevB.79.224509,PhysRevB.84.214509,PhysRevB.85.174532,PhysRevB.88.144503,PhysRevB.90.220511,PhysRevB.90.134521} that could combine with the present one.
In particular, the fact that sample edges are metallic, as observed by in-plane tunneling spectroscopy \cite{PhysRevLett.107.077003}, was shown to generate a large reduction in predicted edge currents\cite{PhysRevB.90.134521}.
Following Ref.~\cite{PhysRevB.90.134521}, we model the metallic edge by a region of width $L_M$ sites where the gap is set to zero.  
As shown in Fig.\ref{BdG}, the presence of a metallic edge generates an even larger drop of the current over a temperature scale given by $T/T_c \sim \xi/L_M$, with $\xi$ the coherence length.

Experimental data\cite{PhysRevB.76.014526,PhysRevB.81.214501,PhysRevB.89.144504} restricts edge currents to be three orders of magnitude smaller than the Matsumoto-Sigrist prediction obained for $d_{z,0}$ \cite{1999JPSJ...68..994M}, which is of the same order as the value we find for $I_0$ at $T=0$.
%
As seen in Fig.~\ref{BdG}, the current predicted for $d_{z,\text{RG}}$ at the temperature relevant to experiments ($T/T_c=0.2$) is roughly three orders of magnitude smaller than $I_0$ at $T=0$.
This prediction could therefore potentially explain the absence of measurable edge currents generated fields.


Admittedly, the weak-coupling RG technique we used to predict $d_{z,\text{RG}}$  is exact only in the $U/t \rightarrow 0$ limit, while this ratio is finite for a realistic material.
The gap in the real material will therefore be renormalized compared to the gap function we find from the RG. 
Nevertheless, the gap function $d_{z,\text{RG}}$ was shown to reproduce the specific heat data\cite{PhysRevB.89.220510}.
Furthermore, $d_{z,\text{RG}}$ has deep minima on $\Balpha$ and $\Bbeta$, as required by STM\cite{PhysRevB.88.134521}(the gap function on $\Bgamma$ cannot be observed directly in STM because of atomic orbitals anisotropy).
Also, finite coupling RG calculations have shown similar results: the pairing on $\Bgamma$ was shown to have a substantial $g_{2}$  component from a singular-mode functional RG calculation\cite{0295-5075-104-1-17013} and a large $g_{3}$ component was shown to be favored from a calculation combining RG with the constrained random phase approximation\cite{2015arXiv150301352T}.


In conclusion, we have shown from a microscopic calculation that a chiral state whose two dominant gap functions are $\sin(3 k_x) + i \sin(3 k_y)$ and $\sin(k_x) \cos(k_y) + i \sin(k_y) \cos(k_x)$ is favored on the three bands of \SROdot, at least in the weak coupling limit.
This OP leads to a Chern number of $-7$, in contrast to the previously assumed value of $+1$.
%
This state naturally predicts both time-reversal symmetry breaking and the possibility of a large reduction of edge currents, thereby helping to reconcile two sets of experiments: optical Kerr effect and muon spin relaxation on one side, and negative results obtained in the search for edge currents on the other.
The present results could be an important piece of the puzzle in reconciling the absence of edge currents with the presence of a chiral superconducting state in \SROdot.

\begin{acknowledgements}
Helpful conversations with Andy Mackenzie, Clifford Hicks, Catherine Kallin, James Sauls, Suk Bum Chung, Masahisa Tsuchiizu and Titus Neupert are acknowledged. This work is supported by EPSRC Grant Nos. EP/I032487/1 and EP/I031014/1, the Clarendon Fund Scholarship, the Merton College Domus and Prize Scholarships, and the University of Oxford.
EPSRC requirements on data management: this publication reports theoretical work that does not require supporting research data.
\end{acknowledgements}

\bibliography{EdgeCurrents}

\end{document}